\documentclass[aps,prl,showpacs,groupedaddress,twocolumn,preprintnumbers,amsmath,amssymb]{revtex4}

\usepackage{graphicx}

\begin{document}

\title{Generation of a spin-squeezed state with trapped ions using a dressing field}

\author{Atsushi~Noguchi$^{1}$}
\email[]{noguchi@qe.ee.es.osaka-u.ac.jp}
\author{Kenji~Toyoda$^{1}$}
\author{Shinji~Urabe$^{1}$}
\affiliation{%
$^{1}$Graduate School of Engineering Science, Osaka University, 1-3 Machikaneyama, Toyonaka, Osaka, Japan}

\date{\today}
             
\begin{abstract}
We propose a method for generating a spin-squeezed state that is a symmetric Dicke state, with trapped ions using only global access.
The eigenstates of the ions under a strong dressing field become symmetric Dicke states and the M$\o$lmer--S$\o$rensen interaction selectively couples one of them to an initially populated auxiliary state.
A $\mid\! D_{2n}^n\rangle$ state, which is  maximally spin squeezed, can be generated with high fidelity using only square pulses. 
Using an adiabatic technique, the ideal maximally spin-squeezed state is generated.
\end{abstract}

\maketitle

The generation of genuine multipartite entanglement is of central importance for quantum information science and quantum information processing. 
Entangled states are resources for quantum computation and can also be used to improve the sensitivity beyond the shot noise limit of an atomic clock or in other precise measurements\cite{1}. In particular, Greenberger--Horne--Zeilinger states and the symmetric Dicke states have been extensively investigated\cite{1,2}.
The symmetric Dicke states are given by
\begin{equation*}
\mid\! D_{n}^m \rangle =\frac{1}{\sqrt{{}_nC_m}}\sum _k \hat{P}_k\mid \uparrow\uparrow\dots \uparrow\downarrow\downarrow\dots\downarrow\rangle, 
\end{equation*}
where $P_k$ are permutation operators and ${}_nC_m =n!/[m! (N-m)!]$.
Among the symmetric Dicke states, {\it{the maximally spin-squeezed state}} ($\mid\!\! D_{2n}^n \rangle$) has the highest sensitivity for precise measurement and achieves the Heisenberg limit\cite{2}.
High-sensitivity measurements using entangled states have been demonstrated by a single pair \cite{3} and two pairs\cite{4,5} of photons and a single pair of trapped ions\cite{6}.
Furthermore, precise measurement using the large-scale entanglement of twin matter waves has recently been reported\cite{7,8}, though its sensitivity has not reached the Heisenberg limit.
The generation of large-scale entanglement with high fidelity remains a significant challenge.

Currently, large-scale entanglement is generated experimentally by using a common phonon mode with trapped ions.
In this method, a phonon mode is excited with either an array of laser pulses irradiated locally\cite{9,10} or by off-resonant global irradiation (geometric phase gate\cite{11} and M$\o$lmer--S$\o$rensen gate\cite{12}).
Using local pulse arrays, a W state ($\mid\! D_n^1\rangle$) of eight ions is generated\cite{10}. Using off-resonant lasers, the generation of a Greenberger--Horne--Zeilinger state of 14 ions has been demonstrated\cite{13}.
As the number of atoms used is increased, local access becomes more difficult and complicated.
It is then important that the generation of entangled states does not require local access.

The generation of a symmetric Dicke state was proposed\cite{14} and demonstrated\cite{15} using an adiabatic technique.
However, this proposal requires a phononic Fock state for the initial state. Also, 
to generate the maximally spin-squeezed state, a large Fock state is needed, which is a challenge for achieving high fidelity.

In this paper, we propose a method for generating a maximally spin-squeezed state using only global irradiation by a laser, radio-frequency or microwave field. 
The combination of a dressing field and the M$\o$lmer--S$\o$rensen interaction\cite{16} has been reported experimentally\cite{17} for two ions. This work presents a theoretical extension to a large number of ions. The method can generate the large-scale symmetric Dicke state, especially the maximally spin-squeezed state ($\mid\! D_{2n}^{n}\rangle$).

\begin{figure}[b]
   \includegraphics[width=7cm,angle=0]{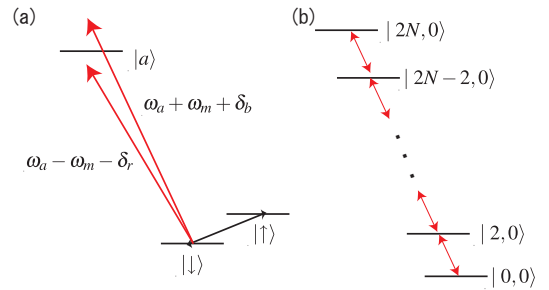}
\caption{(a) Energy-level diagram of trapped $2N$ ions. $\omega _a$ is the frequency of the transition $\lvert a\rangle\leftrightarrow\lvert\downarrow\rangle$, $\omega _m$ is the normal mode frequency and $\delta _{r,b}$ is the amount of detuning from the sideband transition.
(b) Whole Hamiltonian approximated as a single multi-level ladder.
}
\label{model}
\end{figure}

We first consider the situation depicted in fig.\ 1(a).
2N ions are trapped and cooled to the ground state in a linear RF trap whose normal mode frequency of the center of mass mode is $\omega _m$.
There are three long-lived states, namely an ancillary state, $\lvert a\rangle$, and two qubit states, $\{\lvert\downarrow\rangle,\lvert\uparrow\rangle\}$.
The interaction between these states can be expressed by the following Hamiltonian in the interaction picture,
\begin{eqnarray*}
\hat{H}_I &=&\hat{H}_1+\hat{H}_2,\\
\hat{H}_1&=&\sum _{i}\frac{\hbar\Omega _1}{2} (\hat{\sigma}^{(i)+}_{1}+\mathrm{h.c.}),\\
\hat{H}_2&=&\sum _{i}\frac{\hbar\eta \Omega _2}{2} (\hat{a}\hat{\sigma} ^{(i)+}_{2} e^{-i \delta _r t} +\hat{a}^\dagger \hat{\sigma} ^{(i)+}_{2} e^{i\delta _b t} +\mathrm{h.c.}),
\end{eqnarray*}
where $\hat{a}$ and $\hat{a}^\dagger$ are the annihilation and creation operators of the motional mode, respectively, and $\hat{\sigma}^{(i)+}_{1}$ and $\hat{\sigma}^{(i)-}_{1}$ ($\hat{\sigma}^{(i)+}_{2}$ and $\hat{\sigma}^{(i)-}_{2}$) are the spin flip operators between the $\{\lvert\downarrow\rangle , \lvert\uparrow\rangle\}$ ($\{\lvert a\rangle , \lvert\downarrow\rangle\}$) states of the i-th ion. 
The first term ($\hat{H}_1$) gives rise to dressed states of the qubit,  expressed as $\lvert +\rangle=(\lvert\uparrow\rangle+\lvert\downarrow\rangle)/\sqrt{2}$ and $\lvert -\rangle=(\lvert\uparrow\rangle-\lvert\downarrow\rangle)/\sqrt{2}$. $\Omega _1$ is the Rabi frequency of the transition between the qubit states.
The second term in the Hamiltonian ($\hat{H}_2$) corresponds to an MS type interaction\cite{17}.
$\eta \Omega _2$ is the Rabi frequency of the sideband transitions and $\delta _{r,b}$ ($\delta _r =\omega _a-\omega _m-\omega _{lr}, \delta _b =\omega _{lb}-\omega _a - \omega _m$ where $\omega _{lr,lb}$ are the frequency of lasers) is the amount of detuning from the sideband transitions.
To make the MS interaction an effective Ising interaction, the detunings $\delta _{r,b}$ are much larger than the Rabi frequencies $\eta \Omega _2$ 
and the effective Ising interaction is at resonance when $\delta _r=\delta _b =\delta$\cite{17}.

We introduce the global spin operators $\hat{S}_\alpha=\sum _{i}\hat{\sigma}^{(i)}_{\alpha, 1}$ and $\hat{J}_\alpha=\sum _{i}\hat{\sigma}^{(i)}_{\alpha, 2}$,
where $\hat{\sigma}^{(i)}_{\alpha, j}$ $(\alpha =\{x,y,z\})$ is the projection along the $\alpha$ axis of the Pauli matrices.
Using this notation, the Hamiltonian is rewritten such that\cite{17}
\begin{eqnarray*}
\hat{H}_I =\frac{\hbar\Omega _1}{2}\hat{S}_x+\frac{\hbar (\eta\Omega _2)^2}{4\delta}\hat{J}_x^2.
\end{eqnarray*}
The global spin operator $\hat{S}_x$ commutes with the operator $\hat{\textbf{S}}^2$ ($=\hat{S}_x^2+\hat{S}_y^2+\hat{S}_z^2$) and its eigen-states, $\mid\!\! S, m\rangle$, namely the dressed states, can be expressed by the eigenvalues of $\hat{\textbf{S}}^2$ and $\hat{S}_x$.
Because $\hat{J}_x$ and $\hat{S}_x$ are global spin operators and the Hamiltonian commutes with the permutation operators of ions $\hat{P}_k$, the symmetry of the whole system with regard to permutation of ions is preserved during the time evolution with the Hamiltonian.
We consider an initial state $\mid\! aa\dots a\rangle$. Then, the unique completely symmetric state, which does not change with arbitrary exchanges of ions, can couple to this initial state due to the symmetry of the initial state.
We define the state such that
\begin{equation*}
\mid \!\! N_a, m\rangle = \hat{P}_{\mathrm{sym}}[\mid\! a,a,\dots ,a\rangle\otimes\mid\! S_{\mathrm{max}}, m\rangle ],
\end{equation*}
where $N_a$ is the number of ions in the $\mid\! a\rangle$ state and $S_{\text{max}}$ (=$(2N-N_a)/2$) is the maximum value of $S$ when $N_a$ ions are in the $\mid\! a\rangle$ state and $\hat{P}_{\mathrm{sym}}$ is the symmetrizing operator, which is the normalized summation of the operators proportional to the permutations.
Note that the $\mid\!\! N_a,m\rangle$ states include the symmetric Dicke states ($\mid\! D_{2S_{\text{max}}}^{S_{\text{}max}+m}\rangle$) with $2N-N_a$ ions.

We expand the Hamiltonian with these $\mid \!\! N_a, m\rangle$ states:
\begin{eqnarray*}
\hat{H}_I &=&\sum _{N_a, m}\frac{\hbar\Omega _1}{2}m\mid \! N_a, m\rangle\langle N_a, m\!\mid +\frac{\hbar (\eta\Omega _2)^2}{4\delta}\hat{J}_x^2.
\end{eqnarray*}
The first term is diagonalized with these states and we examine the coupling of the states due to the second term.
The initial state ($\mid\!\!\! 2N, 0\rangle$) is not affected by the dressing field and hence this initial state is an eigenstate of the first term of the Hamiltonian with eigenvalue 0.
If $\Omega _1\gg \frac{(\eta\Omega _2)^2}{2\delta}$ is satisfied, the initial state can be coupled only to the $\mid\! N_a, 0\rangle$ states by the Hamiltonian because the other states have large energy due to the dressing field and are greatly detuned from the resonance of the effective Ising interaction. 
The error in this approximation is of the order of $[\frac{(\eta\Omega _2)^2}{2\delta}/\Omega _1]^2$.
Then, the whole Hamiltonian can be factorized into a subspace $\it{D}$ spanned by ${\it{D}}=\mathrm{Span}\{\mid\! N_a, 0\rangle \mid N_a=0,1,\dots , 2N\}$, which includes the initial state and the maximally spin-squeezed state $\mid\! 0,0\rangle$, and the orthogonal complement $\it{D}^\perp$:
\begin{equation*}
\hat{H}_I =
\begin{pmatrix}
\hat{H}_{{\it{D}}}&0\\
0&\hat{H}_{{\it{D}}^\perp}
\end{pmatrix},
\end{equation*}
\begin{equation*}
\hat{H}_{{\it{D}}}=\frac{\hbar (\eta\Omega _2)^2}{4\delta}\hat{J}_x^2.
\end{equation*}
We focus on the subspace ${\it{D}}$ and calculate the matrix elements of this effective Hamiltonian. 
Because $\hat{J}_x^2$ is constructed by the products of two spin flip operators, the matrix elements of $\hat{H}_{{\it{D}}}$ are   
\begin{eqnarray*}
&&\langle N_a, 0\mid \hat{H}_{{\it{D}}}\mid\! N_a^\prime ,0\rangle \\
&=&H_{N_a,N_a^\prime}
\begin{cases}
\neq 0 & (N_a=N_a^\prime , N_a^\prime \pm 2)\\
=0 & (N_a\neq N_a^\prime , N_a^\prime \pm 2).
\end{cases}
\end{eqnarray*}
Then, the interaction on $\it{D}$ can be expressed as a single ladder (fig.\ 1(b)).
The dimension of the effective Hamiltonian reduces to N+1.  

To calculate exact values of the matrix elements, we expand the maximally spin-squeezed state $\mid\!\! S_{\text{max}}, 0\rangle$ with the symmetric Dicke states along the z axis ($\mid \! S, S_z =m_z\rangle$):
\begin{equation*}
\mid\!\! S_{\text{max}}, 0\rangle =\sum _{m_z=-S_{\text{max}}}^{S_{\text{max}}} p(S_{\text{max}},m_z)
\mid \! S_{\text{max}}, S_z =m_z\rangle ,
\end{equation*} 
\begin{equation*}
p(n,m)=\frac{1}{2^n}
\sqrt{ \frac{ {}_{2n}C_{n} }{ {}_{2n}C_{n-m}} }
\sum _{i=0}^{n-m}
(-1)^{n-m-i} 
{}_n C_{n-m-i} \cdot {}_n C_i.
\end{equation*}
Using this expansion, the Hamiltonian is expressed such that
\begin{eqnarray*}
&&\lefteqn{H_{N_a,N_a}}\\ 
&=&\sum _{j=0}^{S_{\text{max}}} \mid p(S_{\text{max}},S_{\text{max}}-2j)\mid ^2 V ^{(2N-2j)}_{S_{\text{max}}-2j,S_{\text{max}}-2j},
\end{eqnarray*}

\begin{eqnarray*}
&&\lefteqn{H_{N_a,N_a- 2}}\\ 
&=&\sum _{j=0}^{S_{\text{max}}-1} \mid\! p(S_{\text{max}},S_{\text{max}}-2j)p(S_{\text{max}}-2,S_{\text{max}}-2j-2)\!\mid \\
&{}&\times V ^{(2N-2j)}_{S_{\text{max}}-2j-2,S_{\text{max}}-2j},
\end{eqnarray*}
where $V^{(n)}_{l,k}$ are the matrix elements of the effective Ising interaction,
\begin{equation*}
V^{(n)}_{l,k}=\langle J=n/2, j_z=l\mid \hat{J}_x\hat{J}_x\mid J=n/2, j_z= k\rangle .
\end{equation*}

\begin{figure}[t]
   \includegraphics[width=7cm]{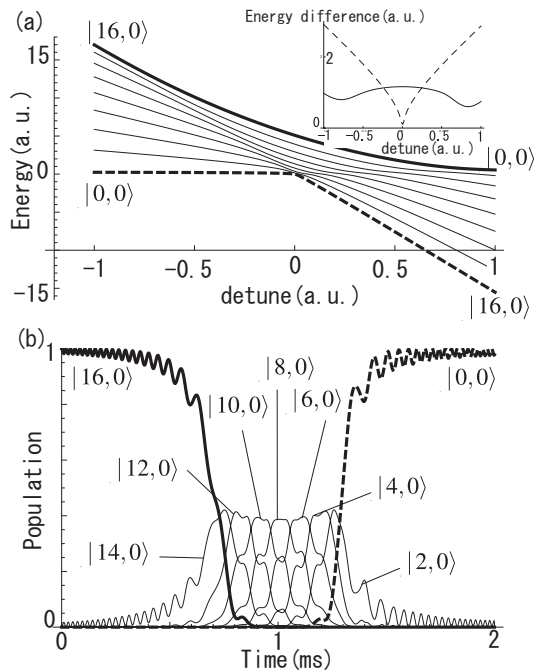}
\caption{(a) Eigenvalues of the Hamiltonian $\hat{H}_{{\it{D}}}$ as functions of the detuning of the effective Ising interaction with 16 ions.
In the inset, the solid and dashed curves shows the energy difference between the two highest and lowest energy states, respectively, which is related to the adiabatic condition.
(b) Generation of the maximally spin-squeezed state by an adiabatic process using 16 ions.
The calculation parameters are as follows: sweep time: 2 ms, detuning linear sweep: from $-2\pi\times 28$kHz to $2\pi\times +28$ kHz, peak strength of the effective Ising interaction: $\frac{(\eta\Omega _2)^2}{2\delta}=2\pi\times 3$ kHz, and amplitude sweep: Gaussian envelope with $e^{-1}$ full width of 1.3 ms.}
\label{adiabatic}
\end{figure}

We numerically analyze the time evolution using this Hamiltonian.
To generate the large-scale maximally spin-squeezed states more effectively, an adiabatic process is useful, as in similar multi-level systems (fig.\ 1(b))\cite{15,18,19,20,21}.
This corresponds to the rapid adiabatic passage (RAP) technique\cite{22} using the effective Ising interaction.
RAP is a robust population-transfer method using a pulse with a time dependent envelope and frequency.
Fig.\ 2(a) shows numerically calculated eigenvalues of the Hamiltonian $\hat{H}_{{\it{D}}}$ against the detuning of the effective Ising interaction ($\Delta = \mid\delta _r\mid-\mid\delta _b\mid )$).
We prepared the $\mid\! 2N, 0\rangle$ state, which corresponds to the curve at the top of fig.\ 2(a), and swept the detuning.
The scalable maximally spin squeezed state $\mid\! 0, 0\rangle$ ($=\mid \mathrm{D}_{2N}^N\rangle$) can be ideally generated through RAP.
The numerical result is shown in fig.\ 2(b) with 16 ions and the parameters used in the calculation are given in the caption.

In ordinary two-level RAP we can use either of two directions for the detuning sweep and both directions are equivalent with regard to population transfer.
However, we here use eigenstates of a non-linear Hamiltonian and hence the two sweep directions are not equivalent.
At the resonance condition ($\Delta$ $=$ 0), the energy difference between the two lowest energy states and that between the two highest energy states are not equal, so that the situation is different depending on the direction of the detuning sweep and on the state that is used (the highest energy state or the lowest energy state).
The inset of fig.\ 2(a) shows the energy difference between the two lowest energy states and two highest energy states of fig.\ 2(a).
The conditions of adiabaticity differ depending on the direction of the detuning sweep due to the non-linearity of the Hamiltonian and more efficient population transfer is achieved with the highest energy state.
Generally, the adiabatic condition is expressed as $\Delta _{ad}\gg \tau ^{-1}$m where $\Delta _{ad}$ is the minimal energy separation to the other states and $\tau$ is the typical time for the detuning and amplitude sweep.
For the 16-ion case (fig.\ 2) the minimal energy separation from the highest energy state is calculated numerically to be $\Delta_{ad}\cong 8.02 (\eta \Omega _{2\mathrm{max}})^2/2\delta$.

To date, RAP has not been demonstrated for the effective Ising interaction, and there are a few technical challenges in applying RAP to the effective Ising interaction.
First, because the effective Ising interaction arises from a two photon process\cite{21}, the strength of the interaction is limited to a few kHz in a conventional RF Paul trap and a long interaction time is required for RAP.
Such a long interaction time induces a large decoherence due to magnetic fluctuations and other disturbances.
Second, to form the maximally spin-squeezed states there are two limiting conditions regarding detuning: $\Omega _1\gg \frac{(\eta\Omega _2)^2}{2\delta}$ and $\delta _{r,b}\gg \eta\Omega _2$. During the detuning sweep these conditions must be satisfied at all time.

\begin{figure}[t]
   \includegraphics[width=9cm,angle=0]{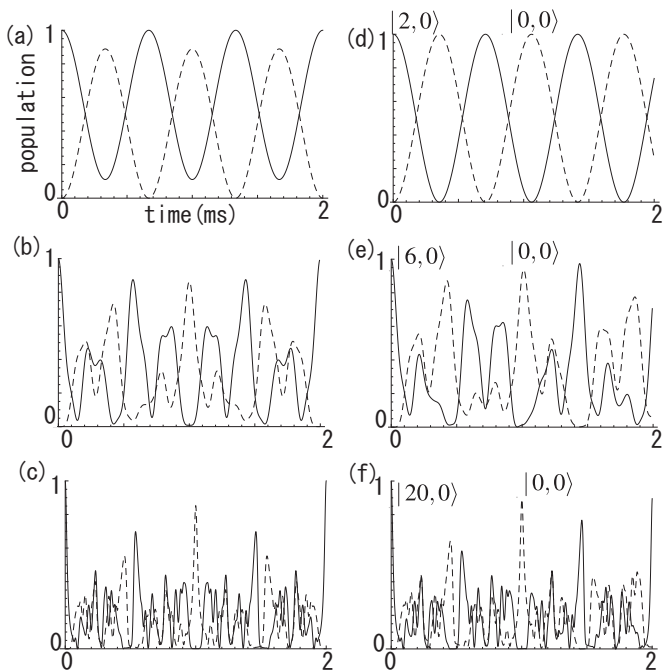}
\caption{Time evolution with square pulses.
(a)--(c) Evolution of the state populations on resonance: (a) 2 ions, (b) 6 ions, (c) 20 ions.
(d)--(f) Evolution with optimized detuning of the effective Ising interaction: (d) 2 ions, (e) 6 ions, (f) 20 ions. 
Each solid and dashed curve expresses the population of the $\mid N_{\mathrm{ion}}, 0\rangle$ states and the $\mid 0, 0\rangle$ states, respectively.
}
\label{setup}
\end{figure}

\begin{figure}[t]
   \includegraphics[width=7cm,angle=0]{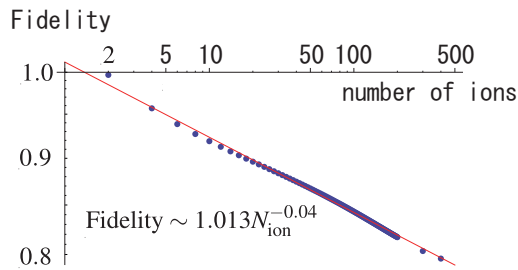}
\caption{Fidelity vs. the number of ions on a log--log plot. A high fidelity ($>0.8$) can be achieved even for 300 ions. 
The decay of the fidelity is approximately calculated as $\text{Fidelity}\sim 0.1013N_{\text{ion}}^{-0.04}$ by fitting these numerical results.}
\label{scalability}
\end{figure}

Although we can generate multipartite maximally spin-squeezed states through an adiabatic process such as RAP, it is technically difficult to use the effective Ising interaction in an adiabatic process. Here we analyze the generation with a single detuning and square pulse. 
For the two-ion case, the Hamiltonian is expressed in the basis of $\{\mid\! 0,0\rangle , \mid \! 2,0\rangle\}$ as
\begin{equation*}
\hat{H}_{{\it{D}}}=\frac{\hbar (\eta\Omega _2)^2}{4\delta}
\begin{pmatrix}
2&\sqrt{2}\\
\sqrt{2}&1
\end{pmatrix}.
\end{equation*}
Note that the diagonal elements are different to each other and a shift appears which non-linearly depends on the number of ions in the $\mid \! a\rangle$ states.
For the two-ion case, we can cancel this shift by adjusting the detuning of the effective Ising interaction $\Delta$, and the ideal maximally spin-squeeze state can be generated.
The time evolution with this Hamiltonian is depicted in fig.\ 3(a) and (d) without and with optimized detuning, respectively.
In other cases, although the non-linear energy shifts are never canceled and become complicated, 
the population of maximally spin-squeezed state can be improved with optimized detuning [fig.\ 3(b), (c), (e), (f)] of the square pulse.

We numerically determined the optimized detuning [$\Delta_{\text{opt}}(N_{\text{ion}})$] that maximizes the fidelity, depending on the ion number:
\begin{equation*}
\Delta _{\text{opt}}(N_{\text{ion}})\cong 0.225\frac{(\eta\Omega _2)^2}{2\delta}N_{\text{ion}}^{-0.7}. 
\end{equation*}
The maximal fidelity with varying number of ions with optimized detuning of the square pulse is shown in fig.\ 4.
By fitting the numerical results, the decrease in fidelity is found to be in proportion to the $-$0.04-th power of the number of ions and a fidelity of more than 0.9 is achieved for 16 ions.

In summary, we propose and analyze the generation of the maximally spin-squeezed state $\mid\! D_{2n}^n\rangle$, which is a symmetric Dicke state and can be used to achieve Heisenberg limited precise measurements, using only trapped ions with global access of the M$\o$lmer--S$\o$rensen (effective Ising) interaction and the dressing field. 
Under the condition that the dressing field is much stronger than the effective Ising interaction, the whole Hamiltonian can be divided into a small subspace and we can calculate the time evolution of the system for a large number of ions.
The fidelity of this state can be made unity by introducing an adiabatic process. 
The ideal large-scale maximally spin squeezed state can be generated by sweeping the detuning of the effective Ising interaction.
The effective Hamiltonian includes non-linear energy shifts.
However, when these shifts are canceled by the detuning of the effective Ising interaction, the maximally spin-squeezed state can be generated with high fidelity, which decreases in proportion to the $-$0.04-th power of the number of ions, even using square pulses.

This work was supported by the MEXT Kakenhi "Quantum Cybernetics" Project and the JSPS through its FIRST Program.
One of the authors (N. A.) was supported in part by the Japan Society for the Promotion of Science.

\end{document}